\documentclass[graybox]{svmult}

\usepackage{mathptmx}       
\usepackage{helvet}         
\usepackage{courier}        
\usepackage{type1cm}        
\usepackage{makeidx}        
\usepackage{graphicx}       
\usepackage{multicol}       
\usepackage[bottom]{footmisc}
\usepackage{tikz}
\usepackage{url}
\usepackage{amsfonts}
\usepackage{amsmath}
\usepackage{amssymb}
\usepackage{xcolor}
\usepackage{algorithm}
\usepackage{algpseudocode}
\usepackage{csquotes}
\usepackage{xspace}

\newcommand{\MRAS}{\texttt{MRASLive}\xspace}

\begin{document}

\title{\MRAS: An online parameter estimation toolbox for time-dependent PDEs}
\author{Christian Aarset, Benedikt Kaltenbach and Tram Thi Ngoc Nguyen}
\institute{C. Aarset \at Aix-Marseille School of Economics, Aix-Marseille University, France
\and 
B. Kaltenbach \at Institute for Numerical and Applied Mathematics, University of G\"ottingen, Germany
\and 
T. T. N. Nguyen \at Department of Applied Mathematics and Theoretical Physics, University of Cambridge, United Kingdom
\email{ttn35@cam.ac.uk}}

\maketitle

\abstract{
We present \MRAS, a toolbox for online parameter identification in time-dependent partial differential equations based on a model reference adaptive system (MRAS). Building on the MRAS proposed by Kaltenbacher and Nguyen (Inverse Problems 37(5) 2021) and its first numerical realization by the authors (Applied Mathematics for Modern Challenges 8 2026), the toolbox is designed for ease of use, utilising the user's weak form specification to prepare the data, the discretized MRAS and its solvers, and to verify the assumptions of MRAS's convergence theory semi-automatically.
}


\section{Introduction}
\label{sec:introduction}

\emph{Data assimilation} is the process of estimating the evolving state of a dynamical system by combining observational data with a model prediction in an \emph{online} fashion \cite{ReichCotter15}. Sequential schemes such as the ensemble Kalman filter \cite{SchillingsStuart17} update the system state whenever new observations become available, while variational methods \cite{Lorenc86} minimize a cost functional that penalizes deviations between estimated states and observations. Although data assimilation is predominantly used for state estimation, it also provides a natural framework for {estimating uncertain parameters} of the model. In contrast to conventional parameter identification \cite{nguyen19,nguyen24bilevel, nguyen25sequential}, the inversion is then carried out simultaneously with data acquisition, continuously refining the parameter estimate as new data arrive.

Model reference adaptive systems (MRAS) form a class of adaptive control schemes \cite{IoannouSun96,NarendraAnnaswamy05} that design dynamic update laws for both parameter and state, driving the estimated output toward the reference trajectory: the state equation is modified by feedback terms, and the parameter evolution is driven by the model-observation mismatch. MRAS approaches for partial differential equations \cite{BaumeisterEtAl97,BoigerKaltenbacher16,Kugler08} were developed mainly for models depending linearly on the unknown parameter. In \cite{KaltenbacherNguyen21}, this restriction was lifted, allowing the model to be nonlinear both in the state and in the parameter. Its first numerical realization, for four benchmark problems of increasing complexity, was given in \cite{KaltenbachAarsetNguyen26}.

In this contribution, we present \MRAS, a toolbox publicly available at \url{https://github.com/CAarset/MRASLive} that makes the MRAS of \cite{KaltenbacherNguyen21,KaltenbachAarsetNguyen26} accessible for general parabolic equations, including vectorial equations that were not examined in \cite{KaltenbachAarsetNguyen26}. A problem is specified by its physics alone; the data, the discretized MRAS and its solvers are derived from it. Moreover, the assumptions of the convergence theory are checked semi-automatically on the problem at hand.

\section{Data assimilation via model reference adaptation}
\label{sec:mras}

\subsection{The parametric equation and data}
\label{sec:equation}

In this section, we recall the setting in \cite{KaltenbacherNguyen21,KaltenbachAarsetNguyen26}. Given a spatial domain $\Omega\subset\mathbb{R}^d$, let $H$ be a Hilbert space of functions on $\Omega$ and let $U\hookrightarrow H\hookrightarrow U^*$ and $Q\hookrightarrow H\hookrightarrow Q^*$ be Gelfand triples of Sobolev spaces, $Q\subseteq U$. We consider the parabolic equation
\begin{equation}\label{eq:pde}
	D_t u^\dagger(t) + f\big(q^\dagger, u^\dagger(t)\big) = g(t),
	\quad t>0, \qquad u^\dagger(0) = u_0,
\end{equation}
with time-dependent state $u^\dagger$, a spatially dependent ground truth parameter $q^\dagger\in Q$ to be determined, a source $g\in L^2([0,\infty);U^*)$ and an initial condition $u_0\in H$. The model
$f : L^2([0, \infty); H) \times L^2([0, \infty); U )\to L^2([0, \infty); U^*)$  is a Nemytskii operator \cite{troeltzsch10}. It captures advection, diffusion and reaction, and is in general nonlinear in both $q$ and $u$. Non-homogeneous, possibly time-dependent Dirichlet data are incorporated by a lifting argument. Since the state and the data carry the same Dirichlet values, the errors $u-z$ and $u-u^\dagger$ vanish on $\Gamma_D$. We assume that \eqref{eq:pde} is uniquely solvable for $u^\dagger$ at $q^\dagger$. The state is observed on $\Omega$ as
\begin{equation}\label{eq:data}
	z(t) \approx u^\dagger(t) \qquad \text{for } t>0 \text{ until the
	current time.}
\end{equation}

If the whole data $z$ were available at once, estimating $q^\dagger$ would be an \emph{inverse parameter  problem}. In data assimilation, $z(t)$ is available only up to the current time. Thus, $q^\dagger$ is instead approximated \emph{online}, i.e.~as more data become available, via a parameter approximation $q(t)$ and a state approximation $u(t)$, in the sense that
\begin{equation}\label{eq:task}
	\lim_{t\to\infty}\|q(t)-q^\dagger\|_H = 0
	\quad\text{and}\quad
	\lim_{t\to\infty}\|u(t)-u^\dagger(t)\|_H = 0.
\end{equation}

\subsection{Online estimation via model reference adaptive system}
In \cite{KaltenbacherNguyen21}, we proposed the model reference adaptive system (MRAS)
\begin{equation}\label{eq:mras}
	\begin{aligned}
	D_t q + \sigma\big[D_t z + f(q,z) - g\big]
		- f_q'(\tilde q, z)^*(u-z) &= 0, \qquad \sigma\in\{0,1\},\\
	D_t u + f(q,z) + \mathcal{C}(\|q\|_H)(u-z) &= g,\\
	(q,u)(0) &= (q_0,u_0)
	\end{aligned}
\end{equation}
for the task of online parameter estimation. The switching parameter $\sigma$ is zero if $f$ is (affine) linear in $q$ and one otherwise. $\tilde q$ is a reference parameter at which $f$ is G\^ateaux differentiable, $q_0$ is an initial guess for $q^\dagger$, and $\mathcal{C}(\|q\|_H)\in\mathcal{L}(U,U^*)$ is a tracking operator. Note that if $\sigma = 0$, $z=u$ and $q=q^\dagger$, the MRAS \eqref{eq:mras} reduces to \eqref{eq:pde}.

In \eqref{eq:mras}, the parameter equation is driven by the model-data residual and the observation residual $u-z$, while the state equation is modified by a feedback term in $u-z$. It is worth noting that \eqref{eq:mras} is linear in $u$ even where $f$ is nonlinear in the state. Convergence of \eqref{eq:mras} rests on the following assumptions \cite[Assumption 2.2]{KaltenbachAarsetNguyen26}:
\begin{description}[(A3)]
	\item[(A1)] There is a monotonically increasing $L:[0,\infty)\to[0,\infty)$ such that, for all $q\in H$ and a.e. in time,
	\[
		\|f(q,z)-f(q^\dagger,z)-f_q'(\tilde q,z)(q-q^\dagger)\|_{U^*} \le
		L(\|q\|_H)\,\|q-q^\dagger\|_H .
	\]
	\item[(A2)] There is $C_{\mathrm{coe}}>0$ such that, for all $q\in Q$ and a.e. in time,
	\[
		\langle f(q,z)-f(q^\dagger,z),\,q-q^\dagger\rangle_{Q^*,Q} \ge
		C_{\mathrm{coe}}\,\|q-q^\dagger\|_H^2 .
	\]
	\item[(A3)] For some $M>0$ and a monotonically increasing $\tilde N$, for all $q\in H$, $v,w\in U$,
	\[
		\begin{aligned}
		\langle\mathcal{C}(\|q\|_H)v,v\rangle_{U^*,U} &\ge
		\Big(\frac{L(\|q\|_H)^2}{2C_{\mathrm{coe}}}+M\Big)\|v\|_U^2,\\
		\langle\mathcal{C}(\|q\|_H)v,w\rangle_{U^*,U} &\le
		\tilde N(\|q\|_H)\,\|v\|_U\|w\|_U .
		\end{aligned}
	\]
\end{description}
Under (A1)--(A3), the MRAS \eqref{eq:mras} with clean data $z=u^\dagger$ is uniquely solvable, and the errors $\varepsilon_q:=q-q^\dagger$ and $\varepsilon_u:=u-u^\dagger$ decay exponentially \cite[Proposition~2.1]{KaltenbacherNguyen21}:
\begin{equation}\label{eq:decay}
	\|\varepsilon_u(t)\|_H^2+\|\varepsilon_q(t)\|_H^2 \le
	e^{-C't}\big(\|\varepsilon_u(0)\|_H^2+\|\varepsilon_q(0)\|_H^2\big),
\end{equation}
with $C':=\min\{C_{\mathrm{coe}},\,2MC_{U\to H}\}$ and $C_{U\to H}$ the norm of the embedding $U\hookrightarrow H$. Where the coercivity (A2) fails \cite[Remark~4.2]{KaltenbachAarsetNguyen26}, persistent excitation of the data \cite{BaumeisterEtAl97} takes its place.  For noisy data, smoothed data $R_{\mathrm{sp}}z$ and $R_{\mathrm{ti}}z$, via the regularizing operators in space  $R_{\mathrm{sp}}$ and in time $R_{\mathrm{ti}}$, take the place of $z$ in the MRAS \eqref{eq:mras}.

\section{Algorithm and toolbox \MRAS}\label{sec:toolbox}

Discretization of the MRAS and
numerical schemes was the main topic of \cite{KaltenbachAarsetNguyen26}.  Below, we present \MRAS's algorithm (Algorithm \ref{alg:mras}) with automatic verification. The most important user specification is the physics of the problem through the weak form of the model $f$, including the adjoint of its derivative. An explicit example of how \MRAS{} handles this is presented in Section \ref{sec:experiments}.


\begin{algorithm}[ht!]
\caption{MRAS}
\label{alg:mras}
\begin{algorithmic}[1]
\Require Model $f$, forcing $g$, boundary $h$, initial conditions $u_0$, $q_0$, time step $\Delta t$, stopping time $T$, known noise level $\delta$; filter constant $s$ if $\delta>0$
\State assemble temporal and spatial filters $R_{\mathrm{ti}}$, $R_{\mathrm{sp}}$ \Comment{see \eqref{eq:filter} and \eqref{eq:heat-step}} 
\State automatically estimate or falsify assumptions (A1)-(A3) via random sampling
\State the data model starts at $u_0$, with boundary values
	$h(0)$; $q_0\gets P_hq_0$
\State $e_h\gets\|P_hq^\dagger-q^\dagger\|_H/\|q^\dagger\|_H$
	\Comment{approximation floor in $H_h$ (if available)}
\For{$n=0,1,\dots,T/\Delta t-1$}
	\State $u^\dagger_{n+1}\gets$ Newton solve of \eqref{eq:pde} with $u^\dagger_{n+1}=h(t_{n+1})$ on Dirichlet boundary $\Gamma_D$
	\If{$\delta>0$}
		\State $z^\delta_{n+1}\gets u^\dagger_{n+1}+e_{n+1}$, with
			$\|e_{n+1}\|_H=\delta\|u^\dagger_{n+1}\|_H$, $e_{n+1}=0$ on $\Gamma_D$
		\State $\bar z_{n+1}\gets R_{\mathrm{ti}}(z^\delta_{n+1},\bar{z}_n)$ \Comment{time filtering}
		\State $z_{n+1}\gets R_{\mathrm{sp}}\bar z_{n+1}$ \Comment{space filtering}
	\Else
		\State $z_{n+1}\gets u^\dagger_{n+1}$
	\EndIf
	\State $(u_{n+1},q_{n+1})\gets$ step \eqref{eq:mras}
	\If{$\delta=0$ and
		$\|q_{n+1}-q^\dagger\|_H\le1.25\,e_h\|q^\dagger\|_H$}
		\State \textbf{stop} \Comment{alternatively, a discrepancy principle when $\delta>0$}
	\EndIf
\EndFor
\end{algorithmic}
\end{algorithm}

\section{Numerical experiments: Vector Burgers flow in a channel}
\label{sec:experiments}

To demonstrate the sweeping applicability of \MRAS{}, we utilize its numerical framework to identify the viscosity of a vector Burgers flow in a channel with obstacles. We note that vectorial equations were not covered in \cite{KaltenbachAarsetNguyen26}.

\begin{figure}[ht!]
	\includegraphics[width=\textwidth]{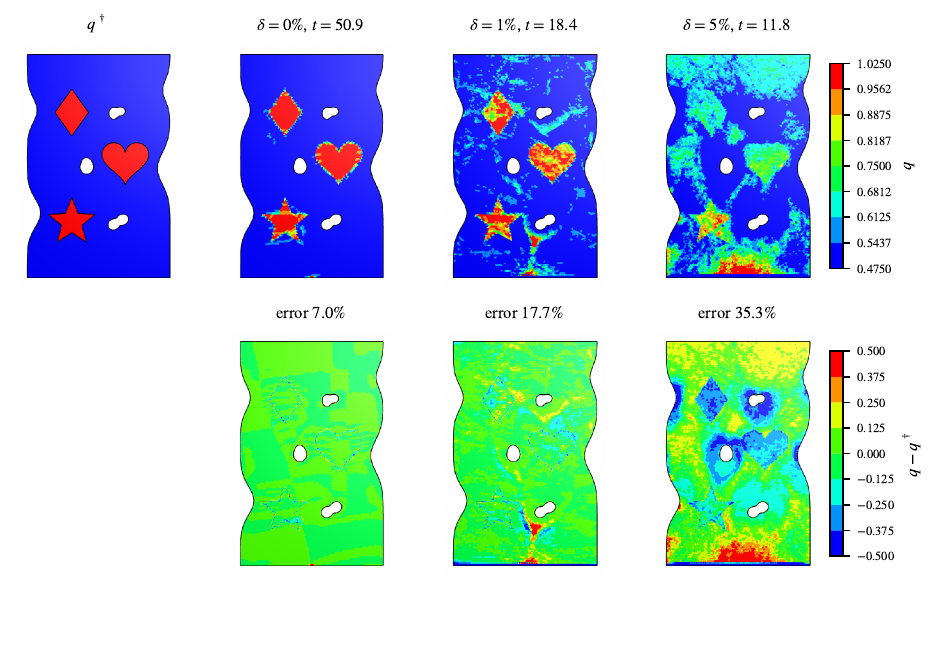}\vspace{-1cm}
	\caption{Top: true viscosity $q^\dagger$ and best iterates $q(t^*)$ for $\delta=0$, $1\%$ and $5\%$. Bottom: Reconstruction errors $q(t^*)-q^\dagger$ and their relative errors.}
	\label{fig:burgers2d-reconstruction}
\end{figure}

\runinhead{Determination of viscosity in Burgers flow.} In a vertical channel $\Omega$ of width $W=3.2$ and height $5$, whose walls bulge inwards at four staggered heights and in which three irregular rounded barriers stand, we identify the viscosity $q$ of the vector Burgers flow
\begin{equation}\label{eq:burgers-2d}
	\begin{aligned}
	D_t u - \nabla\cdot(q\nabla u) + u\cdot\nabla u &= 0
	&& \text{in } (0,\infty)\times\Omega,\\
	u &= h(t,\cdot) && \text{on the inlet } \Gamma_{\mathrm{in}}
		=\{y=0\},\\
	q\,\partial_n u + \beta\,(u\cdot n)\,n &= 0 && \text{on the walls and
		barriers } \Gamma_{\mathrm{s}},\\
	q\,\partial_n u &= 0 && \text{on the outlet } \{y=5\},
	\end{aligned}
\end{equation}
with $u(0)=0$. The above boundary conditions turn walls and barriers into slip surfaces, with no stress along them and with a stiff resistance $\beta=100$ against flow through them. The flow is driven by the non-homogeneous Dirichlet data
\begin{equation}\label{eq:inflow}
	h(t,x) = \rho(t)\,U\Big(\frac14\sin(2\pi t/62.5)\,(1-\xi^2),\ 1-\tfrac12\xi^2\Big),
	\qquad \xi=\frac{2x}{W}-1,
\end{equation}
of centreline speed $U=40$, with smooth ramp $\rho(t)=r^2(3-2r)$, $r=\min\{t/40,1\}$; the first component represents a slowly oscillating transverse flow. The true viscosity $q^\dagger$ is piecewise constant equalling $0.5$ on all of $\Omega$, except taking the value $1$ in a five-pointed star, a heart and a diamond of roughly equal sizes;  see Fig.~\ref{fig:burgers2d-reconstruction}.

\runinhead{MRAS.} With $\langle f(q,u),w\rangle=\int_\Omega q\nabla u:\nabla w +(u\cdot\nabla u)\cdot w\,dx+\langle\mathcal{B}u,w\rangle$ and $\langle\mathcal{B}u,w\rangle=\beta\int_{\Gamma_{\mathrm{s}}}(u\cdot n)(w\cdot n)\,ds$, $f$ is affine in $q$, $\sigma=0$, and $\langle f_q'(\tilde q,z)p,w\rangle=\int_\Omega p\,\nabla z:\nabla w\,dx$, so $f_q'(\tilde q,z)^*w=\nabla z:\nabla w$. Building on the Darcy flow example in \cite[Section~4]{KaltenbachAarsetNguyen26}, we take $\mathcal{C}=-\Delta+\mathcal{B}$. The MRAS for \eqref{eq:burgers-2d} now reads as
\begin{equation}\label{eq:burgers-2d-mras}
\begin{split}
	&D_t q = \nabla z:\nabla(u-z)\\
	&D_t u - \nabla\cdot(q\nabla z) + z\cdot\nabla z - \Delta(u-z)
	+ \mathcal{B}u = 0.
    \end{split}
\end{equation}

\runinhead{Discretization.} Reconstruction is carried out using the library NGSolve \cite{Schoberl14} on a vector $P^3$ space for the state ($\dim U_h=111524$) and a $P^0$ space for the parameter ($\dim H_h=12257$), both on a mesh with maximum element size $h_{\max}=0.05$. The exact viscosity $q^\dagger$ is generated on a mesh of width $h_{\max}/2$ that follows the three shapes, which the estimation mesh does not, thus preventing inverse crime. This yields an approximation floor of $e_h\approx 5.56\%$ relative error in Algorithm \ref{alg:mras}, mostly concentrated along the shape outlines, which no reconstruction can be expected to beat. We moreover set $\Delta t=0.05$, $T=100$ and use initial guess $q_0=3/4$.

\runinhead{Noise.} Noisy data is generated as $z^\delta_n=u^\dagger_n+e_n\delta\|u_n^\dagger\|_H/\|e_n\|_H$, $e_n\sim\mathcal{N}(0,C)$ via the \texttt{ngsrng} package \cite{ngsrng}. Thus, the noise, which is sampled i.i.d.~in each step, is a rescaled coloured centred Gaussian, where $C$ is a Mat\'ern covariance of smoothness $\nu=1$ and correlation length equal to $5\%$ of the domain's diameter.

\runinhead{Smoothing.} $R_{\mathrm{ti}}$ is the causal exponential filter
\begin{equation}\label{eq:filter}
	\bar z_1 = z^\delta_1,\qquad
	\bar z_{n+1} =
	\begin{cases}
	\bar z_n + w\,(z^\delta_{n+1}-\bar z_n) & \text{on the free degrees
		of freedom},\\
	h(t_{n+1}) & \text{on }\Gamma_D,
	\end{cases}
\end{equation}
$w=\Delta t/\tau$, with the time constant $\tau=\max\{\Delta t,\,s\,\delta^{2/3}\Delta t^{1/3}\}$. It leaves a fraction $\delta_f=\delta\,(w/(2-w))^{1/2}$ of noise drawn afresh every step, and lags data varying on a time scale $T_z$ by about $\tau/T_z$. $R_{\mathrm{sp}}$ is one implicit heat step with the boundary values kept,
\begin{equation}\label{eq:heat-step}
	(M_U+\alpha A)\,R_{\mathrm{sp}}\bar z = M_U\,\bar z
	\quad\text{on the free degrees of freedom},
\end{equation}
$A$ is the stiffness matrix, whose $\alpha$ is chosen once, on the first datum that does not vanish, by bisection in $\log\alpha$ for the discrepancy $\|R_{\mathrm{sp}}\bar z-\bar z\|_H=\delta_f\|\bar z\|_H$. The data $z$ is thus everywhere replaced by the smoothed data $z_n:=R_{\mathrm{sp}}(\bar z_n)$. The noisy runs employ the filter constant $s=100$, yielding $\tau=1.7$ and $5.0$ for $\delta=0.01$ and $\delta=0.05$, respectively; the discrepancy rule results in $\alpha=5.0\cdot10^{-7}$ resp.~$1.5\cdot10^{-6}$.

\runinhead{Automatic assumption verification.} \MRAS{}'s automatic verification confirms (A1) with $L=0$ by identifying the affine linear dependence of $f$ on $q$. As $L=0$ and it is easy to verify $\mathcal{B}\geq 0$ via its spectrum, $M=1$ is automatically deduced for Condition (A3), while a lower bound $\tilde N\ge825$ is established numerically. Condition (A2), with the inner product understood via embedding $q$ into two-dimensional vector space, is automatically marked as a failure, random sampling demonstrating that the left-hand side of the condition can take large negative values. As $f$ was verified to be affine linear in $q$, this is sufficient to invalidate the condition regardless of $q^\dagger$. Convergence is therefore attributed to the \emph{persistent excitation} discussed in \cite[Remark~1.6]{KaltenbacherNguyen21}.

\begin{figure}[ht!]
	\includegraphics[width=\textwidth]{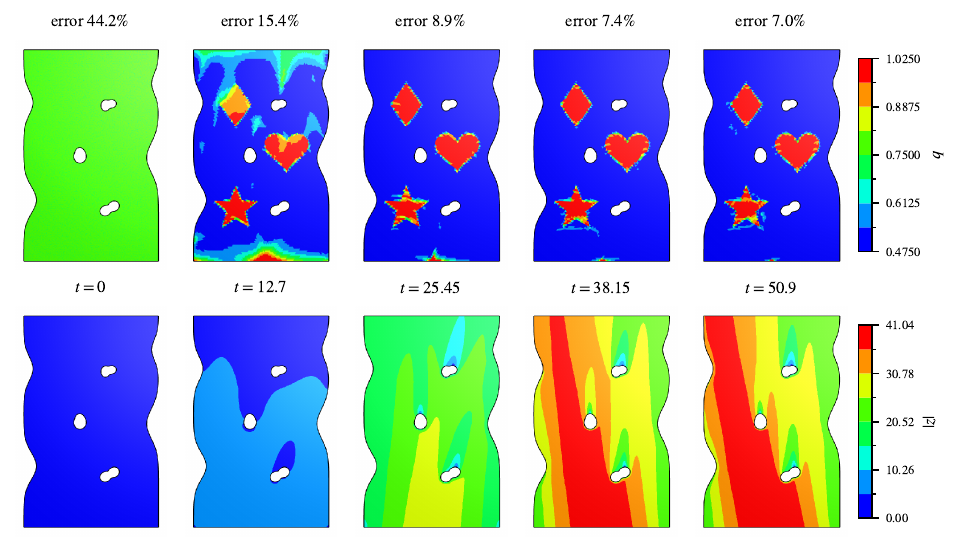}
	\caption{Noise-free reconstruction from initial guess to target at $t=0$, $12.7$, $25.45$, $38.15$ and $50.9$. Top: the estimate $q(t)$ with its relative error $\|q(t)-q^\dagger\|_H/\|q^\dagger\|_H$. Note that error has approached the representation floor $e_h\approx 5.56\%$. Bottom: data velocity $|z(t)|$.}
	\label{fig:burgers2d-snapshots}
\end{figure}

\begin{figure}[ht!]
	\centering
	\includegraphics[width=0.85\textwidth]{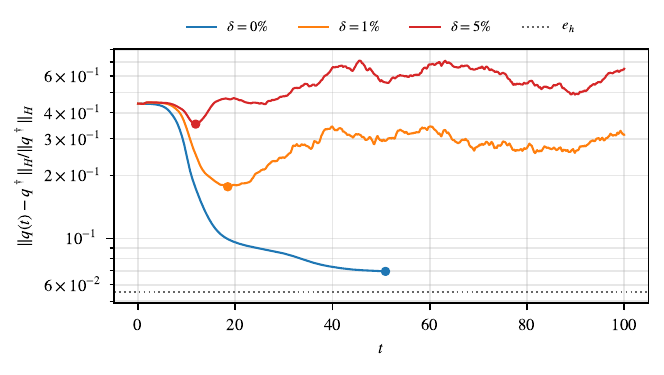}
	\caption{The relative error $\|q(t)-q^\dagger\|_H/\|q^\dagger\|_H$ for $\delta=0$, $1\%$ and $5\%$, the best iterates marked, and the best approximation $e_h=\|P_hq^\dagger-q^\dagger\|_H/\|q^\dagger\|_H$ in $H_h$ (dotted).}
	\label{fig:burgers2d-history}
\end{figure}

\runinhead{Results.} Without noise, the run reaches its target of $1.25\,e_h$ at $t=50.9$, with a relative error of $6.95\%$; parts of the reconstruction trajectory can be seen in Fig.~\ref{fig:burgers2d-snapshots}. With noise, the relative reconstruction error reaches $17.7\%$ at $t=18.4$ for $\delta=1\%$ and $35.3\%$ at $t=11.8$ for $\delta=5\%$, and rises again if the reconstruction is allowed to continue, see Fig.~\ref{fig:burgers2d-history}. Live visualizations of the reconstructions are available at {\small\url{https://github.com/CAarset/MRASLive/tree/main/results/burgers_2d}}.

 \section{Conclusion}
 \label{sec:conclusion}

 We have presented \MRAS, a toolbox for online parameter identification via the model reference adaptive system of \cite{KaltenbacherNguyen21}. With this framework, a problem is specified by its physics via the discretized MRAS and its weak form. \MRAS is computationally efficient, user-friendly and universal, accepting nearly arbitrary weak-form specifications -- including in higher dimensions -- and semi-automatically verifies the required assumptions for convergence analysis.

\begin{acknowledgement}
CA acknowledges support from the Excellence Initiative of Aix–Marseille University - A$^*$MIDEX, a French \enquote{Investissements d’Avenir} programme. TN acknowledges the support of Cambridge Hub for Innovative Mathematics in Research and Application - CHIMiRA, Maths4DL [grant number EP/V026259/1].
\end{acknowledgement}



\begin{thebibliography}{99} 


\bibitem{BaumeisterEtAl97}
Baumeister, J., Scondo, W., Demetriou, M.~A., Rosen, I.~G.: On-line parameter
estimation for infinite-dimensional dynamical systems. SIAM J. Control
Optim. \textbf{35}(2), 678--713 (1997)

\bibitem{BoigerKaltenbacher16}
Boiger, R., Kaltenbacher, B.: An online parameter identification method for
time dependent partial differential equations. Inverse Problems
\textbf{32}(4), 045006 (2016)

\bibitem{IoannouSun96}
Ioannou, P., Sun, J.: Robust Adaptive Control. Prentice Hall, Englewood
Cliffs (1996)

\bibitem{KaltenbacherNguyen21}
Kaltenbacher, B., Nguyen, T.~T.~N.: A model reference adaptive system approach
for nonlinear online parameter identification. Inverse Problems
\textbf{37}(5), 055006 (2021)

\bibitem{KaltenbachAarsetNguyen26}
Kaltenbach, B., Aarset, C., Nguyen, T.~T.~N.: Data assimilation via model
reference adaptation for linear and nonlinear dynamical systems. Applied
Mathematics for Modern Challenges \textbf{8}, 80--112 (2026)


\bibitem{Kugler08}
K\"ugler, P.: Online parameter identification in time-dependent
differential equations as a non-linear inverse problem. European J. Appl.
Math. \textbf{19}(5), 479--506 (2008)


\bibitem{Lorenc86}
Lorenc, A.~C.: Analysis methods for numerical weather prediction. Q. J. R.
Meteorol. Soc. \textbf{112}(474), 1177--1194 (1986)

\bibitem{NarendraAnnaswamy05}
Narendra, K.~S., Annaswamy, A.~M.: Stable Adaptive Systems. Dover, Mineola
(2005)

\bibitem{ReichCotter15}
Reich, S., Cotter, C.: Probabilistic Forecasting and Bayesian Data
Assimilation. Cambridge University Press, Cambridge (2015)

\bibitem{SchillingsStuart17}
Schillings, C., Stuart, A.~M.: Analysis of the ensemble Kalman filter for
inverse problems. SIAM J. Numer. Anal. \textbf{55}(3), 1264--1290 (2017)


\bibitem{ngsrng}
Aarset, C.: 
\texttt{ngsrng}: random functions on finite element spaces. \url{https://github.com/CAarset/ngsrng}

\bibitem{nguyen19}
Nguyen, T.~T.~N.: Landweber--Kaczmarz for parameter identification in time-dependent inverse problems: all-at-once versus reduced version. Inverse Problems \textbf{35}, 035009 (2019)

\bibitem{nguyen24bilevel}
Nguyen, T.~T.~N.: Bi-level iterative regularization for inverse problems in nonlinear PDEs. Inverse Problems \textbf{40}, 045020 (2024)

\bibitem{nguyen25sequential}
Nguyen, T.~T.~N.: Sequential bi-level regularized inversion with application to hidden reaction law discovery. Inverse Problems \textbf{41}, 065015 (2025)

\bibitem{Schoberl14}
Sch\"oberl, J.: C++11 implementation of finite elements in NGSolve. ASC
Report 30/2014, Institute for Analysis and Scientific Computing, Vienna
University of Technology (2014)

\bibitem{troeltzsch10}
Tr{\"o}ltzsch, F.: Optimal Control of Partial Differential Equations: Theory, Methods and Applications. Graduate Studies in Mathematics \textbf{112}. American Math. Soc., Providence, RI (2010)

\end{thebibliography}
\end{document}